\documentclass{article}
\usepackage{amsmath}
\usepackage{spconf,graphicx,bm,amssymb,dsfont,url}
\usepackage{algorithm}
\usepackage{algorithmic}
\usepackage{cases}
\usepackage{nccmath}
\usepackage{cite}
\usepackage{subcaption}
\usepackage{basicNotation}

\newcommand{\fk}{~\forall~k\in\cK}
\newcommand{\fm}{~\forall~m\in\cM}

\title{Joint Beamforming and Compression Design for Per-Antenna Power Constrained Cooperative Cellular Networks}
\makeatletter
\newcommand*\mysize{%
  \@setfontsize\mysize{8.5}{9.0}%
}
\makeatother
\name{Xilai Fan$^{\star,\S}$, Ya-Feng Liu$^{\star}$, and Bo Jiang$^\dagger$}

\address{$^{\star}$LSEC, ICMSEC, AMSS, Chinese Academy of Sciences, Beijing, China \\[2pt]
    $^{\S}$School of Mathematical Sciences, University of Chinese Academy of Sciences, Beijing, China\\[2pt]
  $^\dagger$School of Mathematical Sciences, Nanjing Normal University, Nanjing, China\\[2pt]
Email: fanxilai21@mails.ucas.ac.cn, yafliu@lsec.cc.ac.cn, jiangbo@njnu.edu.cn}

\begin{document}
\ninept

\maketitle

\begin{abstract}
In the cooperative cellular network, relay-like base stations are connected to the central processor (CP) via rate-limited fronthaul links and the joint processing is performed at the CP, which thus can effectively mitigate the multiuser interference. 
In this paper, we consider the joint beamforming and compression problem with per-antenna power constraints in the cooperative cellular network. 
We first establish the equivalence between the considered problem and its semidefinite relaxation (SDR). 
Then we further derive the partial Lagrangian dual of the SDR problem and show that the objective function of the obtained dual problem is differentiable. 
Based on the differentiability, we propose two efficient projected gradient ascent algorithms for solving the dual problem, which are projected exact gradient ascent (PEGA) and projected inexact gradient ascent (PIGA). 
While PEGA is guaranteed to find the global solution of the dual problem (and hence the global solution of the original problem), PIGA is more computationally efficient due to the lower complexity in inexactly computing the gradient. 
Global optimality and high efficiency of the proposed algorithms are demonstrated via numerical experiments. 
\end{abstract}


\section{Introduction}

The cooperative cellular network is a wireless network architecture where multiple relay-like base stations (BSs) are connected to the central processor (CP) through fronthaul links with limited capacities and the joint processing is performed at the CP. 
Therefore, this architecture is able to effectively mitigate the intercell interference among users by enabling multiple BSs to share users' data information via the fronthaul links in cooperatively serving users. 
Nevertheless, the full cooperation among BSs puts heavy burden on the required fronthaul links. 
A promising way of alleviating the stringent requirement on the fronthual links is to jointly design the transmission strategies for the BSs with the utilization of the fronthaul links \cite{peng2015FronthaulconstrainedCloudRadio}. 
Along this direction, various solutions have been proposed \cite{
dai2014SparseBeamformingUsercentric, 
shi2014GroupSparseBeamforming,
park2013JointPrecodingMultivariate, 
park2014InterclusterDesignPrecoding,
patil2014HybridCompressionMessagesharing, 
kang2016FronthaulCompressionPrecoding,
zhou2016FronthaulCompressionTransmit,
he2019HybridPrecoderDesign, 
kim2019JointDesignFronthauling,
ahn2020FronthaulCompressionPrecoding,                                                    
liu2021UplinkdownlinkDualityMultipleaccess,
fan2022EfficientlyGloballySolving,
fan2023QoSbasedBeamformingCompression
}. 
Specifically, the works \cite{liu2021UplinkdownlinkDualityMultipleaccess, fan2022EfficientlyGloballySolving, fan2023QoSbasedBeamformingCompression} considered the joint beamforming and compression problem (JBCP) which minimizes the total transmit power subject to all users' signal-to-interference-and-noise ratio (SINR) constraints and all BSs' fronthual rate constraints. 
In particular, the recent work \cite{fan2022EfficientlyGloballySolving} proposed an efficient fixed point iteration algorithm for solving the JBCP to global optimality, and \cite{fan2023QoSbasedBeamformingCompression} showed the linear convergence rate of the algorithm proposed in \cite{fan2022EfficientlyGloballySolving}. 

In contrast to the prior works that primarily focused on the total transmit power, this paper addresses the more practical per-antenna power constraints (PAPCs) (in the JBCP in the cooperative cellular network). 
First, the PAPC naturally comes from the physical implementation where each antenna has its own power amplifier in its analog front-end, and is limited individually by the linearity of the power amplifier \cite{yu2007TransmitterOptimizationMultiantenna}. 
It is obvious that optimizing/constraining the total transmit power cannot effectively optimize/constrain the power at each antenna (or equivalently at each BS in the cooperative cellular network).  
Second, the existence of PAPCs substantially complicates the solution of the corresponding optimization problems. 
In particular, the optimization problem with and without PAPCs will have different dual problems, and the dual problem of the optimization problem with PAPCs is usually more complicated than that without PAPCs. 
In addition, the existence of PAPCs also makes the efficient duality-based algorithms for solving the corresponding counterpart without PAPCs not directly applicable (because of different problem structures). 
Due to their importance and technical challenges, the beamforming problem with PAPCs has been comprehensively studied under different design objectives and system settings in the literature; see \cite{
yu2007TransmitterOptimizationMultiantenna,
dartmann2013DualityMaxminBeamforming,
zhang2020DeepLearningEnabled,
miretti2023ULDLDualityCellfree,
shi2008PerantennaPowerConstrained,
tolli2008LinearMultiuserMIMO,
christopoulos2014WeightedFairMulticast,
shen2016TransmitterOptimizationPerAntenna,
hu2023SingleLoopAlgorithmWeighted
} and the references therein. 

However, none of the above works studied the JBCP with PAPCs in the cooperative cellular network (possibly because the problem appears to be nonconvex). 
In this paper, we consider the JBCP with PAPCs and propose efficient algorithms for solving the problem based on the recent progress made in \cite{fan2022EfficientlyGloballySolving, fan2023QoSbasedBeamformingCompression} on the JBCP without PAPCs. 
The main contributions of this paper are twofold: 
(1) \emph{Tightness Result}: We establish the tightness of the semidefinite relaxation (SDR) of the considered problem and thus the equivalence between the two problems. 
(2) \emph{Efficient Algorithms}: We further consider solving the problem by deriving and solving the partial Lagrangian dual of the SDR problem. 
We show that the objective function of the dual problem is differentiable. 
Recognizing the differentiability of the dual objective function is of great importance as it facilitates the use of gradient-based algorithms, which is in sharp contrast to slowly convergent subgradient-based algorithms (e.g., \cite{yu2007TransmitterOptimizationMultiantenna, dartmann2013DualityMaxminBeamforming, zhang2020DeepLearningEnabled, miretti2023ULDLDualityCellfree}).
Then we propose two projected gradient ascent algorithms for solving the dual problem, i.e., projected exact gradient ascent (PEGA) and projected inexact gradient ascent (PIGA), where the word ``exact" (``inexact") refers to solving the primal problem (with fixed dual variables) exactly (inexactly) in order to obtain the exact (inexact) gradient of the dual problem. 
Compared to PEGA which is guaranteed to converge to the global solution of the dual problem under mild feasibility assumptions, PIGA is more computationally efficient due to the lower complexity in solving the primal problem in an inexact fashion. 
Numerical experiments show high efficiency and global optimality of PEGA and PIGA.

\emph{Notations.} 
For any matrix $\m{A}$, $\m{A}^\hermitian$ and $\m{A}^\transpose$ denote the conjugate transpose and transpose of $\m{A}$, respectively, and
$\m{A}^{(m, n)}$ denotes the entry on the $m$-th row and the $n$-th column of $\m{A}$. 
We use $\m{0}$ to denote the all-zero matrix of an appropriate size and $\m E_m$ to denote the all-zero square matrix except its $m$-th diagonal entry being one. 
Finally, we use $\cCN(\m{0}, \m{Q})$ to denote the complex Gaussian distribution with zero mean and covariance $\m{Q}$. 

\section{System Model and Problem Formulation}
\subsection{System Model}
Consider a cooperative cellular network consisting of one CP and $M$ single-antenna BSs. 
These BSs are connected to the CP through the noiseless fronthaul links with limited capacities, cooperatively serving $K$ single-antenna users through a noisy wireless channel. 
Let $\cM = \{1,2,\d,M\}$ and $\cK=\{1,2,\d,K\}$ denote the sets of BSs and users, respectively. 

We first introduce the compression model from the CP to the BSs. The beamformed signal at the CP is $ \sum_{k\in\cK} \v{v}_k s_k$, 
where $\v{v}_k = [v_{k,1}, v_{k,2}, \d , v_{k,M}]^\transpose$ is the $M \times 1$ beamforming vector and $s_k\sim \cCN(0, 1)$ is the information signal for user $k$. 
Because of the limited capacities of the fronthaul links, 
the signal from the CP to the BSs need to be first compressed before transmitted. Let the compression error be $\v{e} = [e_1, e_2, \d, e_M]^\transpose \sim \cCN(\m{0}, \m{Q})$, where $e_m$ denotes the error for compressing signals to BS $m$, and $\m{Q}$ is the covariance matrix of the compression noise. 
The transmitted signal of BS $m$ is 
$x_m = \sum_{k\in \cK} v_{k,m} s_k + e_m$ for all $m\in\cM$. 
Then the received signal of user $ k $ is 
$y_k = \sum_{m\in \cM} h_{k,m} x_m + z_k$ for all $k\in\cK$, 
where $h_{k,m}$ is the channel coefficient from BS $m$ to user $k$, and $z_k$ is the additive complex Gaussian noise distributed as $\cCN(0, \sigma_k^2)$.

Under the above model, the received signal at user $k$ is
\begin{equation*}
	y_k = \v{h}_k^\hermitian \left( \sum_{j\in\cK} \v{v}_j s_j \right) + \v{h}_k^\hermitian \v{e} + z_k, \fk, 
	\label{equ:DBM}
\end{equation*}
where $\v{h}_k = [h_{k,1}, h_{k,2}, \d, h_{k,M}]^\hermitian$ is the channel vector of user $k$. 
Then, the transmit power of BS/antenna $m$ is 
$$
\sum_{k\in\cK} |v_{k,m}|^2 + \m Q^{(m,m)},
$$ and the SINR of user $k$ is
\begin{equation*}
	\gamma_k(\{\v v_k\}, \m Q) = \frac{|\v{h}_k^\hermitian \v{v}_k|^2}{\sum_{j\neq k} |\v{h}_k^\hermitian \v{v}_j|^2 + \v{h}_k^\hermitian \m{Q} \v{h}_k + \sigma_k^2},~\forall~k\in\cK. 
\end{equation*}
In order to fully utilize the fronthaul links with limited capacities, we adopt the information-theoretically optimal multivariate compression strategy \cite{park2013JointPrecodingMultivariate} to compress the signals from the CP to the BSs. 
Without loss of generality, we assume that the compression order is from BS $ M $ to BS $ 1 $. Then the fronthaul rate of BS $m$ is given by 
\begin{equation*}\mysize
	\begin{aligned}
		C_m(\{\v v_k\}, \m Q) = \log_2 \frac{\sum_{k\in\cK} |v_{k,m}|^2 + \m{Q}^{(m,m)}}{ \m{Q}^{(m:M, m:M)}/\m{Q}^{(m+1:M, m+1:M)} },\fm. 
	\end{aligned}
\end{equation*}
In the above, $\m{Q}^{(m:M, m:M)}$ denotes the principal submatrix choosing indexes $\{m, m+1, \d, M\}$ and $\m{Q}^{(m:M, m:M)}/\m{Q}^{(m+1:M, m+1:M)}$ denotes the Schur complement of the block $\m{Q}^{(m+1:M, m+1:M)}$ of $\m{Q}^{(m:M, m:M)}$.
\subsection{Problem Formulation}
Given a set of SINR targets $\{\bar\gamma_k\}$, a set of fronthaul capacities $\{\bar C_m\}$, and a set of per-antenna (i.e., per-BS) power budgets $\{\bar P_m\}$, 
we aim to minimize the total transmit power of all BSs subject to all users' SINR constraints, all BSs' fronthaul rate constraints, and all PAPCs:
\begin{equation}
	\begin{aligned}
		\min_{\{\v v_k\}, \m{Q}\succeq \m{0}} &\quad \sum_{k\in\cK} \|\v{v}_k\|^2 + \tr(\m Q)\\
		\st~~~~~ &\quad \gamma_k(\{\v v_k\}, \m Q) \geq \bar\gamma_k, \fk{},\\
		&\quad C_m(\{\v v_k\}, \m Q) \leq \bar C_m, \fm{},\\
		&\quad \sum_{k\in\cK} |v_{k,m}|^2 + \m Q^{(m,m)} \leq \bar P_m, \fm. 
	\end{aligned}
	\label{equ:JBCP_PAPC}
\end{equation}

The last constraint in problem \eqref{equ:JBCP_PAPC} is the PAPC of antenna~$m$ (i.e., BS $m$). 
On the one hand, the PAPC is practically important (as each antenna has its own power budget); on the other hand, PAPCs bring the unique technical difficulty in solving problem \eqref{equ:JBCP_PAPC}. 
More specifically, problem \eqref{equ:JBCP_PAPC} without PAPCs can be solved efficiently and globally via solving two fixed point equations \cite{fan2022EfficientlyGloballySolving}. 
However, the existence of PAPCs makes the algorithm proposed in \cite{fan2022EfficientlyGloballySolving} inapplicable to solve problem \eqref{equ:JBCP_PAPC}. 
In this paper, we shall first reformulate problem \eqref{equ:JBCP_PAPC} as an equivalent semidefinite program (SDP) (in Section~3) and then propose efficient projected gradient ascent algorithms for solving the dual of the equivalent SDP by leveraging the algorithm in \cite{fan2022EfficientlyGloballySolving} (in Section~4). 

\section{Equivalent SDP Reformulation of (1)}
By using the similar arguments as in \cite[Proposition 4]{liu2021UplinkdownlinkDualityMultipleaccess}, we can reformulate problem \eqref{equ:JBCP_PAPC} as problem \eqref{equ:P} at the top of next page. 
\begin{figure*}[!t]
	\begin{equation}
	\label{equ:P}
		\begin{aligned}
			\min_{\{\v v_k\}, \m{Q}\succeq \m{0}}&\quad \sum_{k\in\cK} \|\v{v}_k\|^2 + \tr(\m Q)\\
			\st~~~~~ &\quad  \frac{\bar{\gamma}_k+1}{\bar{\gamma}_k}|\v{v}_k^\hermitian\v h_k|^2 - \sum_{j\in\cK} |\v{v}_j^\hermitian \v h_k|^2 - \v h_k^\hermitian \m Q \v h_k - \sigma_k^2 \geq 0, \fk{}, \\
			&\quad 2^{\bar C_m} \begin{bmatrix}
				\m{0}& \m{0} \\
				\m{0}& \m{Q}^{(m:M, m:M)}
			\end{bmatrix} - \left(\sum_{k\in\cK} |v_{k,m}|^2+ \m{Q}^{(m,m)}\right) \m{E}_m \succeq \m{0}, \fm{}, \\
			&\quad \sum_{k\in \cK} \m |v_{k,m}|^2 + \m Q^{(m,m)} \leq \bar P_m, \fm.
		\end{aligned} 
	\end{equation}
	\hrulefill
\end{figure*}
Problem \eqref{equ:P} is a (non-convex) quadratically constrained quadratic program (QCQP). A well-known technique to tackle the QCQP is the SDR \cite{luo2010SemidefiniteRelaxationQuadratic, xu2023new}. 
Applying the SDR technique to problem \eqref{equ:P}, we obtain problem \eqref{equ:JBCP_PAPC_SDR} at the top of next page, where $\m V_k = \v v_k \v v_k^\hermitian$ for all $k \in \cK$. 
\begin{figure*}[!t]
\begin{subequations}
\label{equ:JBCP_PAPC_SDR}
\begin{align}
	\min_{\{\m V_k \succeq \m 0\}, \m Q \succeq \m 0} &\quad  \sum_{k \in \cK} \tr(\m V_k) + \tr(\m Q) \tag{\ref{equ:JBCP_PAPC_SDR}} \\
	\st~~~~~~~~ &\quad  \frac{\bar\gamma_k+1}{\bar\gamma_k} \tr(\m V_k \v h_k \v h_k^\hermitian) - \sum_{j \in \cK} \tr(\m V_j \v h_k \v h_k^\hermitian) - \tr(\m Q \v h_k \v h_k^\hermitian) - \sigma_k^2 \geq 0, \fk, \label{cst:SINR}\\
	&\quad 2^{\bar C_m} \begin{bmatrix}
		\m 0 & \m 0 \\
		\m 0 & \m Q^{(m:M,m:M)}
		\end{bmatrix} - \left( \sum_{k\in \cK} \m V_k^{(m,m)} + \m Q^{(m,m)} \right) \m E_m \succeq \m 0, \fm, \label{cst:FR}\\
	&\quad \sum_{k\in \cK} \m V_k^{(m,m)} + \m Q^{(m,m)} \leq \bar P_m, \fm. \label{cst:PAPC}
\end{align}
\end{subequations}
\hrulefill
\end{figure*}
We have the following theorem, whose proof is similar to that of \cite[Theorem 1]{fan2022EfficientlyGloballySolving} and is omitted due to the space reason. 
\begin{theorem}
\label{thm:tight}
If problem \eqref{equ:JBCP_PAPC_SDR} is strictly feasible, then its optimal solution $(\{\m V_k^*\}, \m Q^*)$ always satisfies $\rk(\m V_k^*) = 1$ for all $k\in\cK$. 
\end{theorem}

Theorem~\ref{thm:tight} shows that problem \eqref{equ:JBCP_PAPC_SDR} always has a rank-one solution of $\left\{\m V_k^\star \right\}.$ 
Therefore, the SDR of problem \eqref{equ:P} is tight, which means that problem \eqref{equ:JBCP_PAPC_SDR} is an equivalent SDP reformulation of problem \eqref{equ:P}. 
In this case, Theorem~\ref{thm:tight} offers a way of globally solving problem \eqref{equ:P} via solving problem \eqref{equ:JBCP_PAPC_SDR}. 
In the following, instead of directly calling the solver (e.g., CVX \cite{CVX}) to solve the SDP \eqref{equ:JBCP_PAPC_SDR} (due to the high computational cost), we shall design efficient gradient-based algorithms for solving problem \eqref{equ:JBCP_PAPC_SDR} by leveraging the efficient fixed point iteration algorithm in \cite{fan2022EfficientlyGloballySolving}.  

\section{Proposed Gradient-Based Algorithms}
In this section, we propose two gradient-based algorithms for solving problem \eqref{equ:JBCP_PAPC_SDR}. 
Since both of the proposed algorithms are based on solving the dual problem of \eqref{equ:JBCP_PAPC_SDR}, we first derive the dual problem. 
Let $\v \mu = [\mu_1, \mu_2, \d, \mu_M]^\transpose \geq \v 0$ be the Lagrange multiplier associated with the inequality constraints in \eqref{cst:PAPC}. 
Then, after some manipulations, we can obtain the partial Lagrangian dual of problem \eqref{equ:JBCP_PAPC_SDR} as follows:  
\begin{equation}\label{equ:JBCP_PAPC_dual}
    \max_{\v \mu \geq \v 0} \  f(\v \mu) := d(\v \mu) - \sum_{m\in\cM} \mu_m \bar P_m, 
\end{equation}
where $d(\v \mu)$ is the optimal value of the following primal problem:
\begin{equation}
	\label{equ:JBCP_PAPC_dual_sub}
	\begin{aligned}
		\min_{\{\m V_k \succeq \m 0\}, \m{Q}\succeq \m{0}} & \sum_{m \in \cM} (1+\mu_m) \left( \sum_{k\in\cK} \m V_k^{(m,m)} + \m Q^{(m,m)} \right)\\
		\st~~~~~~~~ & \eqref{cst:SINR} \text{~and~} \eqref{cst:FR}. 
	\end{aligned}
\end{equation}

\subsection{Differentiability and Gradient of $f$ in \eqref{equ:JBCP_PAPC_dual}}
According to the classical duality results \cite[p. 216]{boyd2004ConvexOptimization}, the function $f$ in \eqref{equ:JBCP_PAPC_dual} is concave. 
However, it is generally not differentiable. 
Therefore, the subgradient algorithm is often employed to solve the convex problem \eqref{equ:JBCP_PAPC_dual} (e.g., as in \cite{yu2007TransmitterOptimizationMultiantenna, dartmann2013DualityMaxminBeamforming, zhang2020DeepLearningEnabled, miretti2023ULDLDualityCellfree}). 
Fortunately, the following theorem shows that $f$ in our case is differentiable, thereby facilitating the use of efficient gradient-based algorithms in solving problem \eqref{equ:JBCP_PAPC_dual}. 
The differentiability of $f$ is mainly due to the uniqueness of the solution of problem \eqref{equ:JBCP_PAPC_dual_sub} \cite{fan2022EfficientlyGloballySolving}.
\begin{theorem}
\label{lem:diff}
Suppose that problem \eqref{equ:JBCP_PAPC_dual_sub} is strictly feasible. Then $f$ in \eqref{equ:JBCP_PAPC_dual} is differentiable. Moreover, the $m$-th component of the gradient $\nabla f(\v \mu)$ is given by
\begin{equation}
    \label{equ:grad}
     \sum_{k\in\cK} \m V_k^{\star(m,m)}(\v \mu) + \m Q^{\star(m,m)}(\v \mu)  - \bar P_m , \fm, 
\end{equation}
where $\left(\{\m V_k^\star(\v \mu)\}, \m Q^\star(\v \mu\right))$ is the solution to problem \eqref{equ:JBCP_PAPC_dual_sub}. 
\end{theorem}

Theorem~\ref{lem:diff} shows that the gradient $\nabla f(\v \mu)$ at any given point $\v \mu\geq \v 0$ depends on the solution of problem \eqref{equ:JBCP_PAPC_dual_sub}. 
Therefore, to compute the gradient $\nabla f(\v \mu),$ we need to solve problem \eqref{equ:JBCP_PAPC_dual_sub} to global optimality. 
Fortunately, problem \eqref{equ:JBCP_PAPC_dual_sub} is a weighted total transmit power minimization problem subject to all users' SINR constraints and all BSs' fronthaul rate constraints, which can be solved efficiently and globally by the fixed point iteration algorithm proposed in \cite{fan2022EfficientlyGloballySolving}. 
The only difference between problem \eqref{equ:JBCP_PAPC_dual_sub} and the problem considered in \cite{fan2022EfficientlyGloballySolving} is the weights in the objective function. 
However, the weights in problem \eqref{equ:JBCP_PAPC_dual_sub} do not bring any difficulty in solving it by using the algorithm in \cite{fan2022EfficientlyGloballySolving}; see \cite[Algorithm 1]{fan2022EfficientlyGloballySolving} for the detailed description of the algorithm.

\subsection{Proposed PEGA Algorithm}
Since the objective value $f^i := f(\v \mu^i)$ and the gradient $\v g^i := \nabla f(\v \mu^i)$ at the $i$-th iteration can be computed (via solving problem \eqref{equ:JBCP_PAPC_dual_sub}), we are ready to present the projected gradient ascent algorithm for solving problem \eqref{equ:JBCP_PAPC_dual}.
To further accelerate the convergence of the algorithm, we employ the following alternate Barzilai-Borwein (ABB) stepsize \cite{dai2005ProjectedBarzilaiBorweinMethods}: 
\begin{equation}
    \label{equ:BB}
    \alpha^i = \min\left\{\max\left\{\alpha_{\operatorname{ABB}}^i, \alpha_{\min}\right\}, \alpha_{\max}\right\},
\end{equation}
where $0 < \alpha_{\min} < \alpha_{\max}$ are preselected stepsize safeguards, and 
\begin{equation}
\label{equ:alterBB}
    \alpha_{\operatorname{ABB}}^i = 
    \begin{cases}
    \frac{\|\v \mu^{i} - \v \mu^{i-1}\|^2}{\left|(\v \mu^{i} - \v \mu^{i-1})^\transpose (\v g^{i-1} - \v g^{i})\right|}, &\text{if~} i \text{~is~even};\vspace{5pt} \\
    \frac{\left|(\v \mu^{i} - \v \mu^{i-1})^\transpose (\v g^{i-1} - \v g^{i})\right|}{\|\v g^{i-1} - \v g^{i}\|^2}, &\text{otherwise}. 
    \end{cases} 
\end{equation}
Then, the $i$-th iteration of the projected gradient ascent algorithm for solving \eqref{equ:JBCP_PAPC_dual} is given by 
\begin{equation}
\label{equ:GA}
    \v \mu^{i+1} = \left[\v \mu^{i} + \lambda \alpha^{i} \v g^{i}\right]_+,
\end{equation}
where $[\cdot]_+$ denotes the projection operator onto the non-negative orthant and $\lambda > 0$ is a tunable parameter to guarantee the algorithm's convergence.
In particular, we choose $\lambda$ such that $\v \mu^{i+1}$ in \eqref{equ:GA} satisfies the Grippo-Lampariello-Lucidi line search condition \cite{grippo1986NonmonotoneLineSearch, birgin2000NonmonotoneSpectralProjected} with parameters $N \geq 1$ and $\theta \in (0,1)$, given by 
\begin{equation}
\label{equ:GLL}
    f(\v \mu^{i+1}) \geq f_r + \theta {(\v g^{i})}^\transpose (\v \mu^{i+1} - \v \mu^{i}), 
\end{equation}
where $f_r = \min\{f(\v \mu^{i-j}),\ j=0, 1, \d, \min\{N-1, i\}\}$. 
The desirable $\lambda$ that satisfies \eqref{equ:GLL} can be found through the backtracking line search; see Step 2 in Algorithm~\ref{alg:PEGA}. 
The algorithm is terminated if 
\begin{equation}
\label{equ:termination}
     \left\|\left[\v \mu^{i} + \v g^{i}\right]_+ - \v \mu^i\right\| \leq \varepsilon_{\text{out}}, 
\end{equation}
where $\varepsilon_{\text{out}}>0$ is a given error tolerance of solving problem \eqref{equ:JBCP_PAPC_dual}.
This algorithm, named PEGA, is summarized in Algorithm~\ref{alg:PEGA}. 
From the above discussion, we can conclude that PEGA is guaranteed to converge to the optimal solution of problems \eqref{equ:JBCP_PAPC_SDR} and \eqref{equ:JBCP_PAPC_dual} (under mild feasibility assumptions). 
\begin{algorithm}[t]
\caption{Proposed PEGA Algorithm for Solving Problem \eqref{equ:JBCP_PAPC_dual}}\label{alg:PEGA}
\begin{algorithmic}
\STATE Choose $\varepsilon_{\text{out}} > 0$, $N \geq 1$, $\theta \in (0,1)$, $\rho \in (0,1)$, $\v \mu^0 \geq \v 0$, $\alpha^0 > 0$, and $\alpha_{\max}>\alpha_{\min}>0$. 
Apply the fixed point iteration algorithm in \cite{fan2022EfficientlyGloballySolving} to solve problem \eqref{equ:JBCP_PAPC_dual_sub} with $\v \mu=\v \mu^0$ to obtain $f^0$ and $\v g^0$. 
\FOR{$i=0,1,2,\d$}
\STATE \textbf{Step 1}: \emph{(Check termination)} \textbf{if} \eqref{equ:termination} holds \textbf{then} break;
\STATE \textbf{Step 2}: \emph{(Backtracking line search)} Find $\lambda = \rho^j$ such that \eqref{equ:GLL} holds. 
\FOR{$j=0,1,2,\d$}
\STATE Set $\v \mu^{i+1} = \left[\v \mu^i + \rho^j \alpha^i \v g^i\right]_+$. 
\STATE Apply the fixed point iteration algorithm in \cite{fan2022EfficientlyGloballySolving} to solve problem \eqref{equ:JBCP_PAPC_dual_sub} with $\v \mu=\v \mu^{i+1}$ to obtain $f^{i+1}$ and $\v g^{i+1}$. 
\STATE \textbf{if} \eqref{equ:GLL} holds \textbf{then} break;
\ENDFOR
\STATE \textbf{Step 3}: \emph{(Compute ABB stepsize)} Use \eqref{equ:BB} and \eqref{equ:alterBB} to compute $\alpha^{i+1}$. 
\ENDFOR
\end{algorithmic}
\end{algorithm}

\subsection{Proposed PIGA Algorithm}
In the PEGA algorithm, each usage of the objective value $f^i$ and the gradient $\v g^i$ requires solving problem \eqref{equ:JBCP_PAPC_dual_sub} exactly, which is time consuming. 
To further improve the efficiency of PEGA, we propose to solve problem \eqref{equ:JBCP_PAPC_dual_sub} in an inexact but controllable fashion. 
In this way, we can only get some approximations of $f^i$ and $\v g^i$, and we name the corresponding algorithm PIGA. 
To be specific, we terminate the fixed point iteration algorithm in \cite{fan2022EfficientlyGloballySolving} in solving \eqref{equ:JBCP_PAPC_dual_sub} when the violations of the two fixed point equations are less than or equal to the error tolerance $\varepsilon_{\text{in}} > 0$. 
It turns out that this simple criterion can effectively control the error in approximately computing the gradient and at the same time significantly reduce the computational complexity. 
In particular, we choose diminishing $\varepsilon_{\text{in}}$ in our implementation. 

\section{Numerical Experiments}
\begin{figure}
    \centering
    \includegraphics[width=0.45\textwidth]{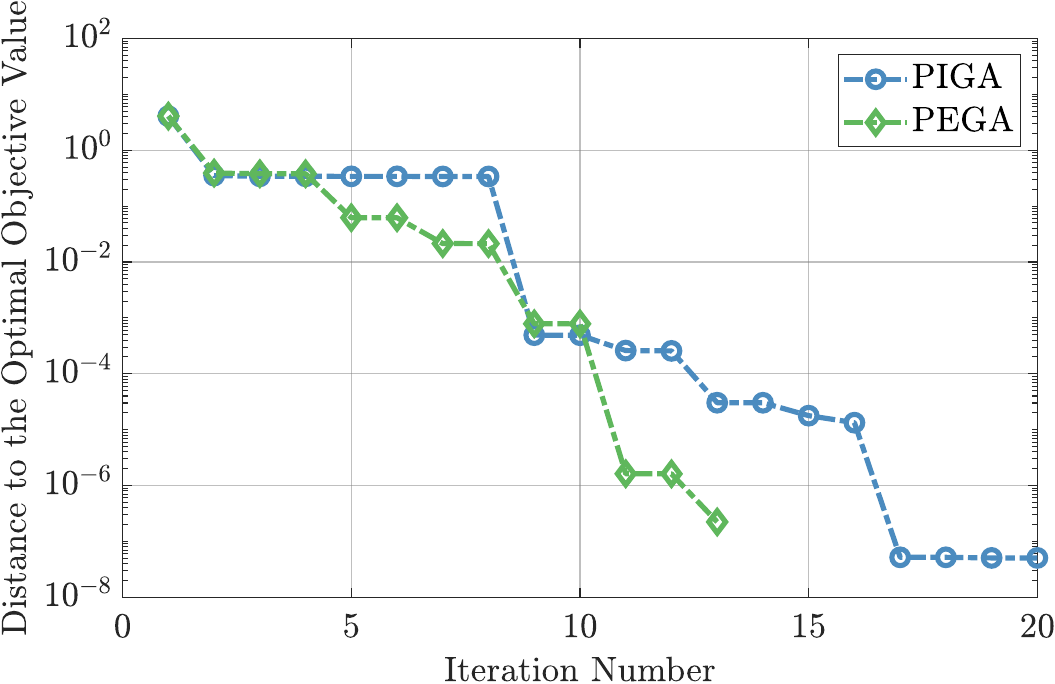}
    \vspace{-8pt}
    \caption{The distance to the optimal objective value versus the iteration number for PEGA and PIGA.}
    \label{fig:1}
    \vspace{-15pt}
\end{figure}
\begin{figure}[!t]
	\centering
	\subfloat[][]{\includegraphics[width=0.22\textwidth]{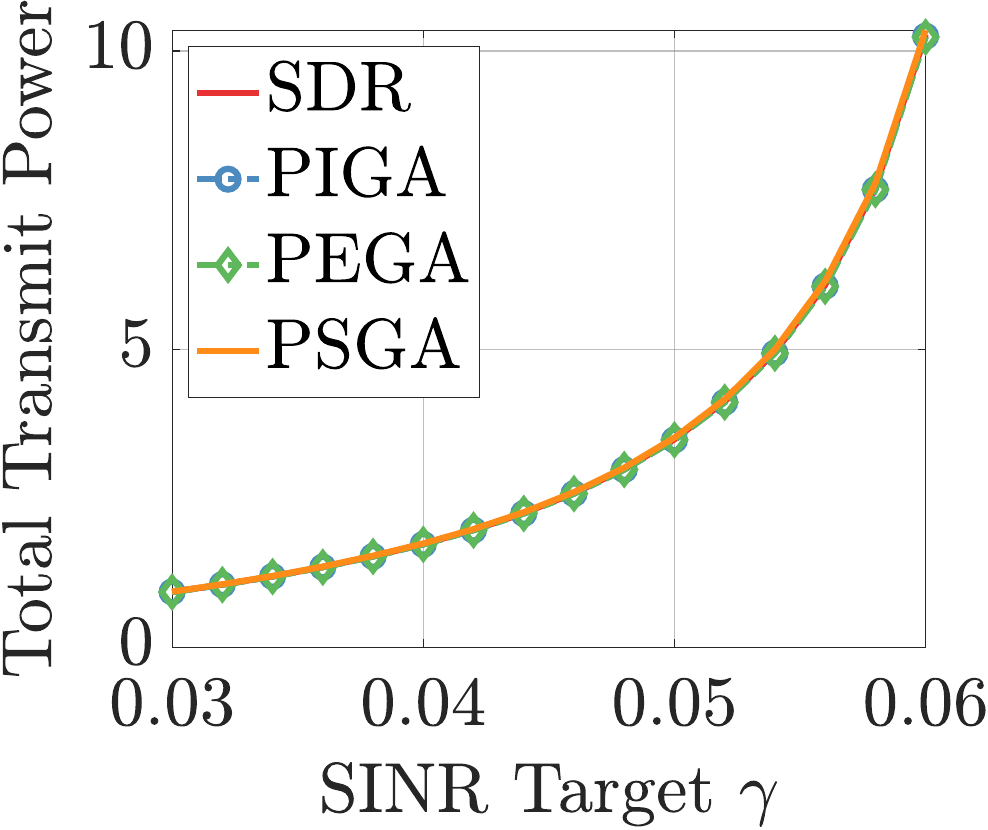}
	\label{fig:2}}
    \subfloat[][]{\includegraphics[width=0.22\textwidth]{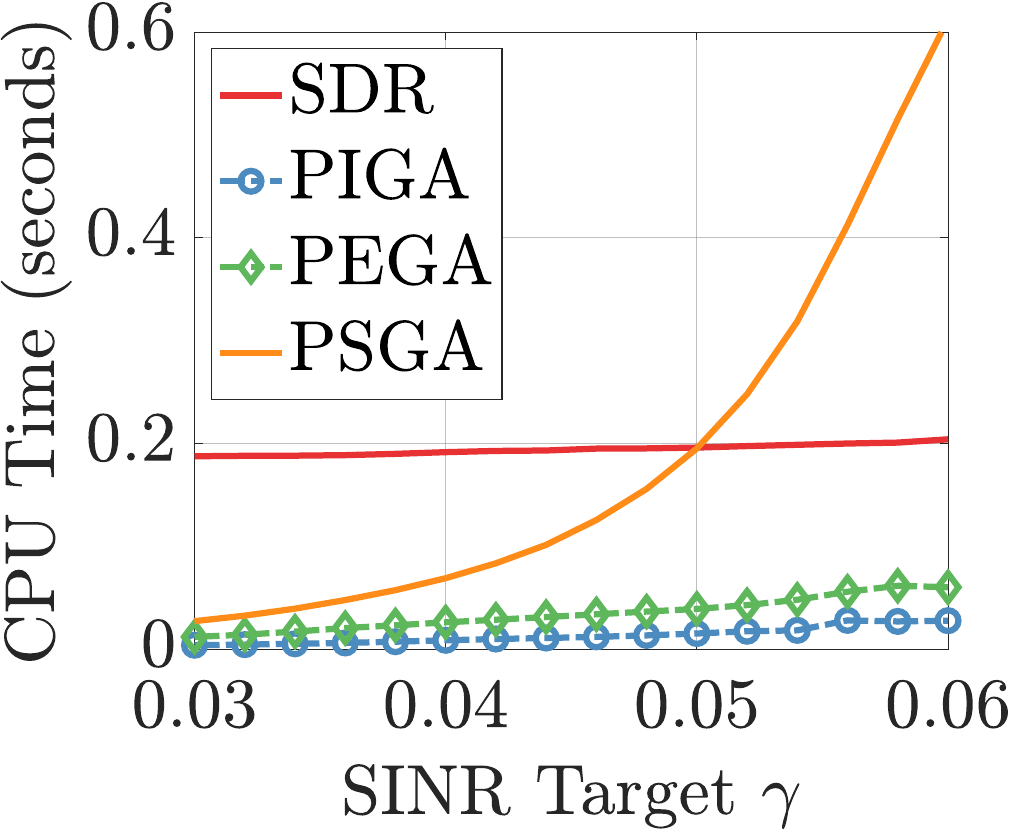}
	\label{fig:3}}\\
    \vspace{-8pt}
	\caption{(a) The total transmit power of the obtained solution versus the SINR target; (b) The average CPU time versus the SINR target. 
	}
	\label{fig:2n3}
	\vspace{-15pt}
\end{figure}

In this section, we present numerical results demonstrating the correctness and efficiency of the proposed algorithms for solving problem \eqref{equ:P}. 
We generate the simulation parameters as in \cite{fan2022EfficientlyGloballySolving, fan2023QoSbasedBeamformingCompression}. 
We consider a network with $M = 8$ and $K = 10$. 
Wireless channels between the BSs and the users are generated based on the i.i.d. Rayleigh fading model following $\mathcal{CN} (0, 1)$. 
The fronthaul capacity $\bar C_m$ between BS $m$ and the CP is set to be $\log_2(1.1)$ for all $m \in \cM$. 
Additionally, the noise power $\sigma_k^2$ at user $k$ is set to be $1$ for all $k \in \cK$. 
The SINR target $\bar\gamma_k$ for user $k$ is set to be $\gamma$ for all $k\in\cK$. 
The per-antenna power budgets are set to $\bar P_m = 8.5$ for $m = 2, 3, \d, M$, and $\bar P_1 =  8.5\times10^{-3}$. 
This setting ensures that at least one PAPC (antenna 1) is active at the optimal solution. 
For both PEGA and PIGA, we set $N=10$, $\theta=10^{-4}$, $\rho = 0.25$, $\v s^0 = \v 0$, $\alpha^0 = 300$, $\alpha_{\min} = 10^{-4}$, and $\alpha_{\max} = 10^{12}$. 
For PIGA, we set $\varepsilon_{\text{in}} = 10^{-3}(i+1)^{-2}$. 

Fig.~\ref{fig:1} compares the convergence behaviour of the two proposed algorithms when applied to solve an instance of problem \eqref{equ:P} with SINR target $\gamma = 0.06$. 
As can be seen from Fig.~\ref{fig:1}, both of the proposed algorithms quickly converge to the optimal objective value (obtained by solving the SDP in \eqref{equ:JBCP_PAPC_SDR}) within $20$ iterations. 
Compared to PIGA, PEGA converges much faster at the cost of a higher per-iteration complexity (in computing the exact gradient). 

In Fig.~\ref{fig:2n3}, we compare our proposed algorithms with the following two important benchmarks. 
\textbf{SDR}: This benchmark directly calls CVX \cite{CVX} to solve problem \eqref{equ:JBCP_PAPC_SDR}. 
The benchmark is helpful in verifying the tightness of problem \eqref{equ:JBCP_PAPC_SDR} as well as the global optimality of PEGA. 
\textbf{PSGA}: The projected subgradient ascent (PSGA) algorithm treats the gradient $\nabla f(\v s)$ as a subgradient, and hence the diminishing stepsize is used and chosen to be $\alpha^{i} = \alpha^0 (i+1)^{-0.1}$ (to guarantee the convergence), where $\alpha^0 = 300$ is chosen for fair comparison. 
This benchmark shows the great importance of recognizing the differentiability of the objective function in \eqref{equ:JBCP_PAPC_dual}. 
The results in Fig.~\ref{fig:2n3} are obtained by averaging over $200$ Monte-Carlo runs, and the termination tolerance for all algorithms is set to $\varepsilon_{\text{out}} = 10^{-3}$. 

Fig.~\ref{fig:2n3}~(\subref{fig:2}) plots the average total transmit power obtained by the four algorithms, where the SINR target $\gamma$ ranges from 0.03 to 0.06. 
This subfigure shows that all of the compared four algorithms return the same global solution. 
This verifies the tightness of the SDR (i.e., Theorem \ref{thm:tight}) and the global optimality of PEGA. 
Although PIGA lacks a theoretical convergence guarantee, Fig.~\ref{fig:2n3}~(\subref{fig:2}) demonstrates that it converges to the same global solution as SDR, PEGA, and PSGA. 
Fig.~\ref{fig:2n3}~(\subref{fig:3}) plots the average CPU time taken by different algorithms. 
It is shown that PEGA and PIGA significantly outperform PSGA in terms of efficiency, and the efficiency advantage quickly increases as the SINR target grows. 
This illustrates the slow convergence of the subgradient algorithm, which in turn shows the importance of recognizing the differentiability of the objective function in \eqref{equ:JBCP_PAPC_dual} (i.e., Theorem \ref{lem:diff}) as well as the use of the adaptive stepsizes in \eqref{equ:BB} and \eqref{equ:alterBB}, as opposed to the diminishing stepsizes used in PSGA. 
Furthermore, PEGA and PIGA also outperform SDR in terms of efficiency, and PIGA exhibits the highest efficiency. 
These preliminary results clearly show global optimality and high efficiency of proposed PEGA and PIGA algorithms.

\newpage
\bibliographystyle{IEEEtran}
\bibliography{cranPerAntenna, misc}

\begin{thebibliography}{10}
\providecommand{\url}[1]{#1}
\csname url@samestyle\endcsname
\providecommand{\newblock}{\relax}
\providecommand{\bibinfo}[2]{#2}
\providecommand{\BIBentrySTDinterwordspacing}{\spaceskip=0pt\relax}
\providecommand{\BIBentryALTinterwordstretchfactor}{4}
\providecommand{\BIBentryALTinterwordspacing}{\spaceskip=\fontdimen2\font plus
\BIBentryALTinterwordstretchfactor\fontdimen3\font minus \fontdimen4\font\relax}
\providecommand{\BIBforeignlanguage}[2]{{%
\expandafter\ifx\csname l@#1\endcsname\relax
\typeout{** WARNING: IEEEtran.bst: No hyphenation pattern has been}%
\typeout{** loaded for the language `#1'. Using the pattern for}%
\typeout{** the default language instead.}%
\else
\language=\csname l@#1\endcsname
\fi
#2}}
\providecommand{\BIBdecl}{\relax}
\BIBdecl

\bibitem{peng2015FronthaulconstrainedCloudRadio}
M.~Peng, C.~Wang, V.~Lau, and H.~V. Poor, ``Fronthaul-constrained cloud radio access networks: Insights and challenges,'' \emph{IEEE Wireless Commun.}, vol.~22, no.~2, pp. 152--160, Apr. 2015.

\bibitem{dai2014SparseBeamformingUsercentric}
B.~Dai and W.~Yu, ``Sparse beamforming and user-centric clustering for downlink cloud radio access network,'' \emph{IEEE Access}, vol.~2, pp. 1326--1339, Oct. 2014.

\bibitem{shi2014GroupSparseBeamforming}
Y.~Shi, J.~Zhang, and K.~B. Letaief, ``Group sparse beamforming for green cloud-{{RAN}},'' \emph{IEEE Trans. Wireless Commun.}, vol.~13, no.~5, pp. 2809--2823, May 2014.

\bibitem{park2013JointPrecodingMultivariate}
\BIBentryALTinterwordspacing
S.-H. Park, O.~Simeone, O.~Sahin, and S.~Shamai, ``\BIBforeignlanguage{en}{Joint precoding and multivariate backhaul compression for the downlink of cloud radio access networks},'' \emph{\BIBforeignlanguage{en}{IEEE Trans. Signal Process.}}, vol.~61, no.~22, pp. 5646--5658, Nov. 2013.
\BIBentrySTDinterwordspacing

\bibitem{park2014InterclusterDesignPrecoding}
------, ``Inter-cluster design of precoding and fronthaul compression for cloud radio access networks,'' \emph{IEEE Wireless Commun. Lett.}, vol.~3, no.~4, pp. 369--372, Aug. 2014.

\bibitem{patil2014HybridCompressionMessagesharing}
P.~Patil and W.~Yu, ``Hybrid compression and message-sharing strategy for the downlink cloud radio-access network,'' in \emph{Proc. {{Inf}}. {{Theory Appl}}. {{Workshop}} ({{ITA}})}, Feb. 2014, pp. 1--6.

\bibitem{kang2016FronthaulCompressionPrecoding}
J.~Kang, O.~Simeone, J.~Kang, and S.~Shamai, ``Fronthaul compression and precoding design for {{C-RANs}} over ergodic fading channels,'' \emph{IEEE Trans. Veh. Technol.}, vol.~65, no.~7, pp. 5022--5032, Jul. 2016.

\bibitem{zhou2016FronthaulCompressionTransmit}
\BIBentryALTinterwordspacing
Y.~Zhou and W.~Yu, ``\BIBforeignlanguage{en}{Fronthaul compression and transmit beamforming optimization for multi-antenna uplink {{C-RAN}}},'' \emph{\BIBforeignlanguage{en}{IEEE Trans. Signal Process.}}, vol.~64, no.~16, pp. 4138--4151, Aug. 2016.
\BIBentrySTDinterwordspacing

\bibitem{he2019HybridPrecoderDesign}
S.~He, Y.~Wu, J.~Ren, Y.~Huang, R.~Schober, and Y.~Zhang, ``Hybrid precoder design for cache-enabled millimeter-wave radio access networks,'' \emph{IEEE Trans. Wireless Commun.}, vol.~18, no.~3, pp. 1707--1722, Mar. 2019.

\bibitem{kim2019JointDesignFronthauling}
J.~Kim, S.-H. Park, O.~Simeone, I.~Lee, and S.~S. Shitz, ``Joint design of fronthauling and hybrid beamforming for downlink {{C-RAN}} systems,'' \emph{IEEE Trans. Commun.}, vol.~67, no.~6, pp. 4423--4434, Jun. 2019.

\bibitem{ahn2020FronthaulCompressionPrecoding}
S.~Ahn, S.-I. Park, J.-Y. Lee, N.~Hur, and J.~Kang, ``Fronthaul compression and precoding optimization for {{NOMA-based}} joint transmission of broadcast and unicast services in {{C-RAN}},'' \emph{IEEE Trans. Broadcast.}, vol.~66, no.~4, pp. 786--799, Dec. 2020.

\bibitem{liu2021UplinkdownlinkDualityMultipleaccess}
\BIBentryALTinterwordspacing
L.~Liu, Y.-F. Liu, P.~Patil, and W.~Yu, ``\BIBforeignlanguage{en}{Uplink-downlink duality between multiple-access and broadcast channels with compressing relays},'' \emph{\BIBforeignlanguage{en}{IEEE Trans. Inf. Theory}}, pp. 7304--7337, Nov. 2021.
\BIBentrySTDinterwordspacing

\bibitem{fan2022EfficientlyGloballySolving}
\BIBentryALTinterwordspacing
X.~Fan, Y.-F. Liu, and L.~Liu, ``\BIBforeignlanguage{en}{Efficiently and globally solving joint beamforming and compression problem in the cooperative cellular network via {{Lagrangian}} duality},'' in \emph{\BIBforeignlanguage{en}{Proc. {{IEEE ICASSP}}}}, May 2022, pp. 5388--5392.
\BIBentrySTDinterwordspacing

\bibitem{fan2023QoSbasedBeamformingCompression}
\BIBentryALTinterwordspacing
X.~Fan, Y.-F. Liu, L.~Liu, and T.-H. Chang, ``{{QoS-based}} beamforming and compression design for cooperative cellular networks via {{Lagrangian}} duality,'' Jun. 2023. [Online]. Available: \url{http://arxiv.org/abs/2306.13962}.
\BIBentrySTDinterwordspacing

\bibitem{yu2007TransmitterOptimizationMultiantenna}
W.~Yu and T.~Lan, ``\BIBforeignlanguage{en}{Transmitter optimization for the multi-antenna downlink with per-antenna power constraints},'' \emph{\BIBforeignlanguage{en}{IEEE Trans. Signal Process.}}, vol.~55, no.~6, pp. 2646--2660, Jun. 2007.

\bibitem{dartmann2013DualityMaxminBeamforming}
G.~Dartmann, X.~Gong, W.~Afzal, and G.~Ascheid, ``On the duality of the max-min beamforming problem with per-antenna and per-antenna-array power constraints,'' \emph{IEEE Trans. Veh. Technol.}, vol.~62, no.~2, pp. 606--619, Feb. 2013.

\bibitem{zhang2020DeepLearningEnabled}
J.~Zhang, W.~Xia, M.~You, G.~Zheng, S.~Lambotharan, and K.-K. Wong, ``Deep learning enabled optimization of downlink beamforming under per-antenna power constraints: {{Algorithms}} and experimental demonstration,'' \emph{IEEE Trans. Wireless Commun.}, vol.~19, no.~6, pp. 3738--3752, Jun. 2020.

\bibitem{miretti2023ULDLDualityCellfree}
\BIBentryALTinterwordspacing
L.~Miretti, R.~L.~G. Cavalcante, E.~Bj{\"o}rnson, and S.~Sta{\'n}czak, ``\BIBforeignlanguage{en}{{{UL-DL}} duality for cell-free massive {{MIMO}} with per-{{AP}} power and information constraints},'' Jan. 2023. [Online]. Available: \url{http://arxiv.org/abs/2301.06520}.
\BIBentrySTDinterwordspacing

\bibitem{shi2008PerantennaPowerConstrained}
S.~Shi, M.~Schubert, and H.~Boche, ``Per-antenna power constrained rate optimization for multiuser {{MIMO}} systems,'' in \emph{Proc. {{International ITG Workshop Smart Antennas}}}, Feb. 2008, pp. 270--277.

\bibitem{tolli2008LinearMultiuserMIMO}
A.~Tolli, M.~Codreanu, and M.~Juntti, ``Linear multiuser {{MIMO}} transceiver design with quality of service and per-antenna power constraints,'' \emph{IEEE Trans. Signal Process.}, vol.~56, no.~7, pp. 3049--3055, Jul. 2008.

\bibitem{christopoulos2014WeightedFairMulticast}
D.~Christopoulos, S.~Chatzinotas, and B.~Ottersten, ``Weighted fair multicast multigroup beamforming under per-antenna power constraints,'' \emph{IEEE Trans. Signal Process.}, vol.~62, no.~19, pp. 5132--5142, Oct. 2014.

\bibitem{shen2016TransmitterOptimizationPerAntenna}
H.~Shen, W.~Xu, A.~L. Swindlehurst, and C.~Zhao, ``Transmitter optimization for per-antenna power constrained multi-antenna downlinks: {{An SLNR}} maximization methodology,'' \emph{IEEE Trans. Signal Process.}, vol.~64, no.~10, pp. 2712--2725, May 2016.

\bibitem{hu2023SingleLoopAlgorithmWeighted}
X.~Hu and X.~Dai, ``A single-loop algorithm for weighted sum rate maximization in multiuser {{MIMO}} systems with per-antenna power constraints,'' \emph{IEEE Trans. Wireless Commun. (early access)}, May 2023.

\bibitem{luo2010SemidefiniteRelaxationQuadratic}
\BIBentryALTinterwordspacing
Z.-Q. Luo, W.-K. Ma, A.~M.-C. So, Y.~Ye, and S.~Zhang, ``\BIBforeignlanguage{en}{Semidefinite relaxation of quadratic optimization problems},'' \emph{\BIBforeignlanguage{en}{IEEE Signal Process. Mag.}}, vol.~27, no.~3, pp. 20--34, May 2010.
\BIBentrySTDinterwordspacing

\bibitem{xu2023new}
Y.~Xu, C.~Lu, Z.~Deng, and Y.-F. Liu, ``New semidefinite relaxations for a class of complex quadratic programming problems,'' \emph{J. Global Optim.}, vol.~87, pp. 255--275, Sep. 2023.

\bibitem{CVX}
M.~Grant and S.~Boyd, ``{CVX}: Matlab software for disciplined convex programming, version 2.1,'' \url{http://cvxr.com/cvx}, Mar. 2014.

\bibitem{boyd2004ConvexOptimization}
S.~Boyd and L.~Vandenberghe, \emph{\BIBforeignlanguage{en}{Convex Optimization}}.\hskip 1em plus 0.5em minus 0.4em\relax {Cambridge University Press}, 2004.

\bibitem{dai2005ProjectedBarzilaiBorweinMethods}
\BIBentryALTinterwordspacing
Y.-H. Dai and R.~Fletcher, ``\BIBforeignlanguage{en}{Projected {{Barzilai-Borwein}} methods for large-scale box-constrained quadratic programming},'' \emph{\BIBforeignlanguage{en}{Numer. Math.}}, vol. 100, no.~1, pp. 21--47, Mar. 2005.
\BIBentrySTDinterwordspacing

\bibitem{grippo1986NonmonotoneLineSearch}
L.~Grippo, F.~Lampariello, and S.~Lucidi, ``A nonmonotone line search technique for {{Newton}}'s method,'' \emph{SIAM J. Numer. Anal.}, vol.~23, no.~4, pp. 707--716, Aug. 1986.

\bibitem{birgin2000NonmonotoneSpectralProjected}
\BIBentryALTinterwordspacing
E.~G. Birgin, J.~M. Mart{\'i}nez, and M.~Raydan, ``Nonmonotone spectral projected gradient methods on convex sets,'' \emph{SIAM J. Optim.}, vol.~10, no.~4, pp. 1196--1211, Jan. 2000.
\BIBentrySTDinterwordspacing

\end{thebibliography}

\end{document}